# A New Weighted Time Window-based Method to Detect B-point in Impedance Cardiogram

Nadica Miljković (ORCiD: https://orcid.org/0000-0002-3933-6076)[*,1,2], Tomislav B. Šekara (ORCiD: https://orcid.org/0000-0001-8031-3135)[1]

1: University of Belgrade – School of Electrical Engineering, Bulevar kralja Aleksandra 73, 11000 Belgrade
2: Faculty of Electrical Engineering, University of Ljubljana, Tržaška 25, 1000 Ljubljana
e-mails: nadica.miljkovic@etf.bg.ac.rs, tomi@etf.bg.ac.rs

**Abstract**

Background: Impedance Cardiogram (ICG) has impressive number of applications in healthcare. However, its wider adoption is excessively limited due to the well recognized challenges in ICG delineation. We present a simple, adaptive, and efficient method for the most demanding ICG delineation task – detection of less distinct B-point that marks the onset of the left ventricular ejection.
Method: The core of the new method is transformation of ICG time series by weighted time window of an ICG segment preceding the maximal ICG peak (the C-point) aiming at the B-point enhancement. The resulting Modified B-point (MB-point) is then easily delineated. To evaluate the proposed workflow, the dataset comprising 20 healthy participants and 21065 B-points are manually annotated and openly shared with the software code. To the best of our knowledge, our ICG dataset has the largest number of annotated B-points. Detector performance is evaluated on two recorded segments with distinct noises, as well as on an open dataset comprising ICG recorded in another set of 24 healthy subjects.
Results: The results showed that the method was superior to the competing methods when the tolerance for B-point detection was set to ±150 ms in all cases and for both datasets (>99.4%).
Conclusions: Proposed approach based on the weighted time windows presents a promising technique for reliable ICG delineation that is instrumental for appropriate feature extraction and ICG characterization. Further customization could incorporate labeling of other pathological ICG segments, other biomedical signals, as well as additional enhancements for improving robustness to noise.

**Abbreviations**
Acc, Accuracy; CO, Cardiac Output; CRAN, The Comprehensive R Archive Network; DE, Detected Error; ECG, Electrocardiogram; FD, Failed Detection; GNU, GNU's not Unix; ICG, Impedance Cardiography; LVET, Left Ventricular Ejection Time; me, mean error; MD, Missed Detection; MB-point, Modified B-point; PPG, Photopletismogram; PP, Positive Predictivity; PEP, Pre-Ejection Period; RMS, Root Mean Square; Se, Sensitivity; sd, standard deviation; SV, Stroke Volume; TP, True Points

---

[*] Corresponding Author: Associate Professor Nadica Miljković, Primary affiliation: University of Belgrade – School of Electrical Engineering, Address: Bulevar kralja Aleksandra 73, 11000 Belgrade, Serbia, e-mail: nadica.miljkovic@etf.bg.ac.rs, Phone: +381 11 3218 348

# 1 Introduction

Impedance Cardiography (ICG) is a cost-effective, simple, portable, and noninvasive technique for monitoring continuous mechanical activity of cardiovascular system that can be used for assessment of heart abnormalities. The technique comprises electrical stimulator that injects alternating current $I$ with high frequency (20 kHz – 100 kHz) and low amplitude (1 mA – 5 mA) through a pair of stimulating electrodes and a system for acquisition of the resulting voltage $U$ with sensing inner pair of surface electrodes. Digitized pulsating signal is proportional to the time-varying bioimpedance ($\Delta U = \Delta Z I$) across the thorax. Commonly, $\Delta Z$ is differentiated and the negative first derivative $dZ/dt \approx \Delta Z/\Delta t$ in [Ω/s] (rate of biompedance change) is known as ICG signal. ICG resembles bioimpedance thorax variation caused by the changes in the blood volume and flow during cardiac cycle, airflow in the lungs, and lung perfusion. [1-11]

ICG is used in astonishing number of applications. Particularly, ICG is used to: obtain Cardiac Output (CO), calculate Stroke Volume (SV), generate impedance distribution of tissue, diagnose heart failure or its cause, assess the severity of heart attack, measure total body water, assess the effect of physical exercises, evaluate sleep disorders, determine the effects of medications on cardiovascular system, predict valvular heart diseases, enable blood flow monitoring during anesthesia, observe hemodynamics in hemodialysis, to monitor thrombosis in the false lumen, etc. Besides, it is used to estimate optimal pacemaker settings, to model volume conductor, and in biometric applications. [1-6, 12-14, 15]

Though it was popularized in 1990s for assessment of systolic time intervals and CO [1], ICG was introduced back in 1932 year by Atzler and Lehmann [16-17]. In years to come, ICG processing methods were developed and explored with the focus on artifact suppression, fiducial points detection, and feature extraction. Despite the outstanding progress, existing methods are still very limited and prevent wider ICG adoption and standardization [1, 4]. Unlike electrocardiogram (ECG), the morphology of ICG is less stationary, making ICG delineation the most challenging part of the analysis. Non-stationary nature of ICG is likely caused by (1) artifacts originating from respiration, motion, and from the poor electrode contacts, and (2) natural ICG variation across subjects and across beat to beat intervals [1, 4, 12]. This subsequently decelerates the application of automatic workflows, as well as error-free detection of cardiovascular indices [4, 6, 18].

Characteristic points of ICG are A-, B-, C- (sometimes denoted as E-point or $(dZ/dt)_{max}$), X-, and O-points. The most prominent C-point is related to the ventricular blood injection and associated with the peak in the aortic blood velocity *i.e.*, ventricular contraction. Contrary to the C-point, the B-point is less distinct. B-point is defined as the notch just before the C-point occurring at the onset of the final ICG rapid upslope. Commonly, B-point is used to determine the onset of left ventricular ejection and matches the beginning of the systole. [1, 3-6, 9-12, 18-19]

The most demanding part of ICG delineation is the B-point detection, mainly as B-point is crucial for reliable and accurate calculation of electromechanical systole intervals such as Pre-Ejection Period (PEP) and Left Ventricular Ejection Time (LVET). PEP is defined as the interval between Q-wave in ECG and B-point, while LVET presents distance between B- and X-points in ICG. Particularly important for psychophysiology studies is PEP interval as it relates to the contractile force of the left ventricle, while LVET is required for SV and CO computation. [1-2, 4-5, 11]

There is no widely accepted standard for B-point location [5]. Current state-of-the-art methods for B-point detection are based on wavelet decomposition, application of higher order derivatives, and on R-peak detection in simultaneously recorded ECG [4]. Commonly, these methods are computationally demanding and require referent ECG recording for accurate delineation which prevents real-time monitoring applications [9]. The proposed approach does not rely on the simultaneously recorded referent ECG signal for ICG delineation albeit it is frequently applied approach [1-3, 6]. We advocate for a new weighted time window-based method to detect B-point with the reference to the prominent C-point [2, 4] as we aim at simple and efficient single signal recordings. At the same time, we propose a beat-to-beat localization of B-point as in [9]. This is not a painless approach as the straight thinking strategy deploys time domain filtering *i.e.*, ensemble averaging to remove artifacts in order to provide a rigorously reliable measures of SV and CO [11].

The new method is inspired by the compelling review in [1] where Authors advocated for boosting the B-point that corresponds to the highest increase in the rising slope before C-point. Rather than employing the second or higher order derivatives [1], which are fairly sensitive to artifacts and measurement noises, we propose a weighted scaling of the upward trend preceding the C-point to enhance the B-point. Our aim is to develop a simple and elegant solution based on the weighted time window scaling for B-peak detection.

## 2 Materials and Methods

For data analysis we use RStudio development environment (Rstudio Inc., Boston, USA) and R programming language v4.1.2 [20] with the following packages: (1) dplyr [21], (2) signal [22], (3) ggplot2 [23], and (4) pracma [24] from the CRAN (The Comprehensive R Archive Network). The analysis was performed on the sample comprising 20 university students (16 females) with no known cardio-vascular disorders (age of 20.80±4.82 years). Although we report participants' sex referring solely to the biological attributes, where information on sex is provided on voluntary bases by the participants themselves, we did not consider any sex differences nor studied them in our paper. ICG data are randomly selected from a larger dataset [25-26]. The primary goal of the original large-scale study was to investigate audio-evoked emotions by physiological measurements. All participants signed Informed Consents in accordance with the Helsinki Declaration and the Institutional Review Board from the Department of Psychology at the University of Belgrade approved the study (No. 2018-19).

### 2.1 Measurement Protocol

For the first two minutes of the recording subjects were instructed to relax in a sitting position (baseline recording). Later, participants were engaged in emotionally induced task. At the end, a post-manipulation period took part. Subjects were not completely restricted from movements after baseline recording. Therefore, we separately analyzed baseline and the complete ICG recording to compare results of B-point detector with and without movement artifacts.

ICG data were recorded with BIOPAC MP160 device and AcqKnowledge software (Biopac Systems Inc., Camino Goleta, CA, USA) with sampling frequency of 2000 Hz. Surface Ag/AgCl electrodes (Kendall, Dublin, Ireland) for ICG stimulation and measurement were placed on the subjects' back following the base of the neck and the xiphoid process as anatomical landmarks. The distance between stimulation and measurement electrodes is kept constant at 3 cm following BIOPAC manual. Injected alternating current frequency is set to 100 kHz and current RMS (Root Mean Square) to 5 mA.

## 2.2 Time Window Scaling Workflow for MB/B-point Detection

The proposed approach for transforming an ICG segment preceding the C-point by weighted time window for B-point enhancement resulting in Modified B-point (MB-point) is given in Fig. 1.

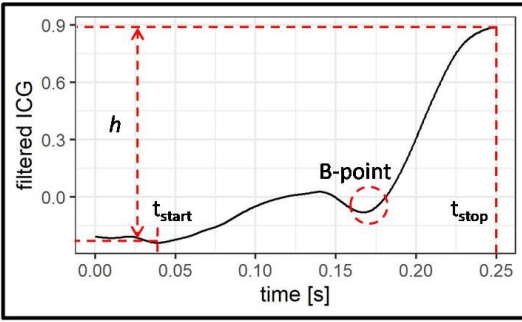
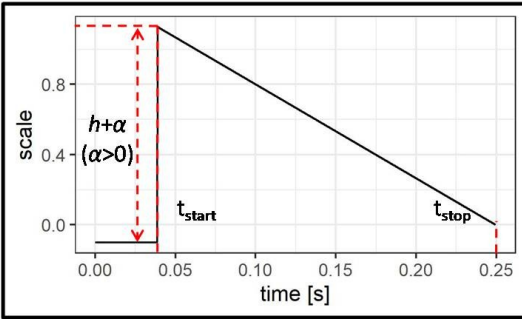
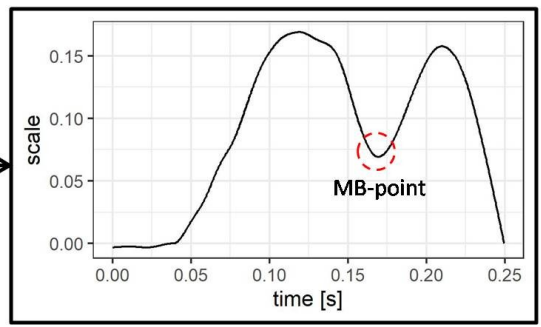

Figure 1, Sample ICG segment comprising B-point (marked with red circle) preceding the C-point (upper-left panel) with weighted time window (lower-left panel) and the resulting modified ICG signal with more distinctive B-point termed Modified B-point (MB-point) marked with red circle (right panel). Segment corresponds to ICG in subject IDN18 and to 20$^{th}$ B-point. Peak-to-peak distance for constructing a weighted window is marked with $h$, while the beginning and end of the scaling window are marked with $t_{start}$ and $t_{stop}$. Parameter $α$ defines a constant scale for constructing weighted time window.

The processing workflow for B-point transformation with weighted time window into MB-point and for MB/B-point detection is presented in Fig. 2. ICG signal is filtered with 3$^{rd}$ order band pass Butterworth filter with cutoff frequencies of 0.5 Hz and 50 Hz (Fig. 2) to remove noise. Zero-phase filtering in forward and backward directions is applied to compensate for phase distortion [7]. After preprocessing, C-points locations are detected as local extremes with simple thresholding method: C-points are determined as local maxima with minimal peak distance of 350 ms and with threshold that equals 80% of standard deviation of ICG signal. The threshold is selected empirically, while the minimal peak distance corresponds to pulse rate of 171.4 beats per minute (bpm).

Then, we construct adaptive weighted time window and apply it to the ICG segment incorporating 250 ms long interval preceding the C-point (upper-left panel in Fig. 1). The segregated segment is afterwards shifted so that the minimal value equals zero. Weighted time window is constructed of negative linear slope from minimal to maximal value of ICG segment. Remaining values were populated with $-\alpha$ ($\alpha$ = 0.1) scale that is set by error and trial with aim to diminish values around B-point. Peak-to-peak of the resulting scaling window is $h + \alpha$ where $h$ presents the difference between maximal and minimal amplitude.

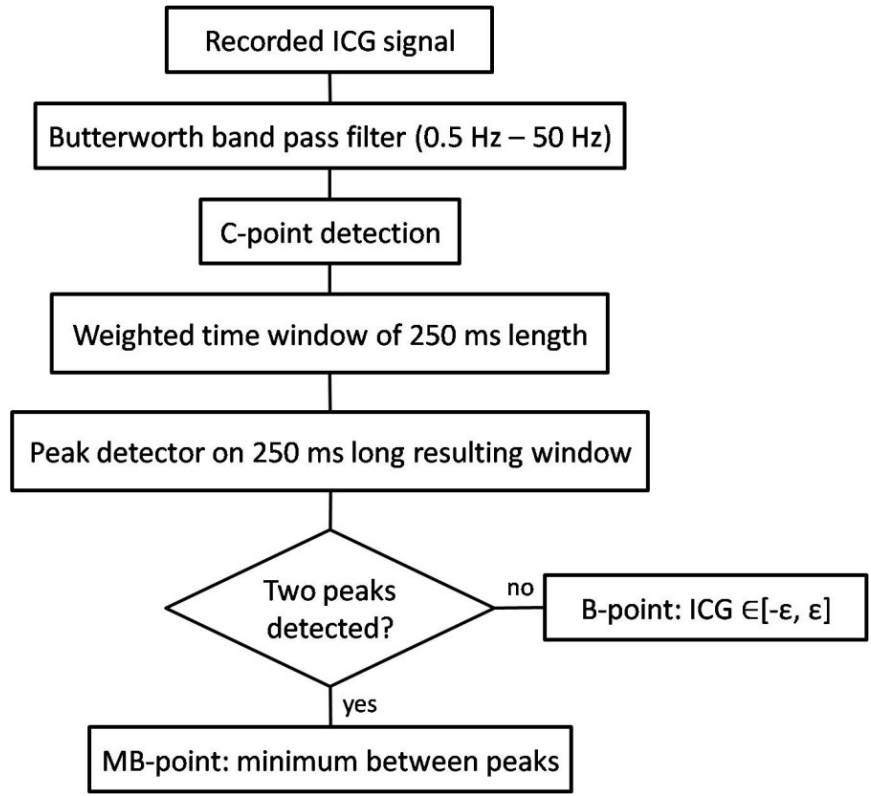

Figure 2, Block diagram of processing workflow for MB/B-point detection from ICG signal by weighted time window method. If scaling window fails to detect two peaks on the resulting window, the B-point is determined as the point in [-ε, ε] (ε > 0 is a small value) in non-weighted ICG segment. Otherwise Modified B-point (MB-point) location is detected as minimum between the two distinctive peaks in transformed ICG.

Next, the scaling was performed and the expected resulting waveform is shown on the right panel in Fig. 1. MB-point is located between two local maxima on a transformed ICG segment. To locate MB-point, we firstly detect the two peaks of a squared and scaled ICG segment with the following criteria: (1) minimal peak distance is set to 50 ms, (2) threshold equals maximal value of current segment divided with 2000, and (3) the number of identified peaks following the first and the second criteria should equal two. Criteria for the peak detector applied on a transformed ICG segment are established empirically and inspired partly by the well-known Pan-Tompkins real-time QRS detector [2-3, 27]. In the event that two peaks are detected MB-point is located as minimum between them. Otherwise, B-point is located in [– ε, ε]. We were inspired by approaches in [1, 4-5] to define B-point location in cases when proposed new algorithm could not distinguish the B-point position. The number of missed MB-point detections that actually correspond to the alternate B-point lineation is reported. The R code with data in 20 healthy subjects is available in [28]. Prior to the publication on Zenodo repository, data have been made anonymous.

## 2.3 Performance Metrics for the MB/B-point Detector

For identifying manual annotations GNU Octave v5.2.0 was used [29]. Firstly, ICG signal is filtered and C-points are detected in the same manner as shown in Fig. 2 and explained previously. Secondly, ICG segments preceding C-point location to 0.5 s after the current C-point location are plotted without marked C-points for proper visibility of B-point incisura. Manual B-points labeling is performed by a moving cursor by the first Author blinded for the automatic B-point locations as in [10]. The procedure was repeated. Lastly, locations of both C- and B-points are saved. B-point is detected for all segments and the judgment for its visual detection is based on the presence of rapid upstroke toward C-point [2]. Additional selective manual inspection of C-point location is performed with satisfactory result. The number of manually detected peaks for each subject is reported. GNU Octave code for manual annotation and saved annotations are freely available in [28]. To ensure well grounded comparison, new method was applied on the available data from [2, 30] with the application of R.matlab library for importing data [31].

For assessment of detector performance, we calculate Accuracy (Acc) as the number of properly detected B-points within tolerance ranges, Sensitivity (Se), Positive Predictivity (PP), Detected Error (DE), and mean error (me) with standard deviation (sd) following previously reported metrics [2, 4, 6, 8, 10, 30]. Moreover, we report comparison with other studies and number of ICG cycles, as well as the subjects (healthy and/or patients). Manual B-points are used for quantitative evaluation as ground truth following [2, 4, 6]. Performance metrics in percents (Se, PP, and DE) are calculated according to the following relations:

$$Se = 100 \times TP/(TP+FD) \quad (1)$$
$$PP = 100 \times TP/(TP+MD) \quad (2)$$
$$DE = 100 \times (FD+MD)/(TP+FD) \quad (3)$$

FD stands for Failed Detection, MD for Missed Detection (detection of a false point), and TP for number of True Points detected. We adopted previously used tolerances of ±30 ms and of ±150 ms from the manually annotated point [2, 10] for appropriate comparison.

## 3 Results

In Tables 1-2, the evaluation metrics for MB/B-point detection in all subjects are given for the complete measurement and for the baseline recording, respectively. PP and Se were not reported as PP equals 100%, while Se is identical to Acc.

In Table 3, comparison of the proposed method for MB/B-point detection with previously published results is presented. In Fig. 3 sample segments for three ICGs with different detection accuracy (IDN18 with Acc150 = 100% and Acc30 = 94.90%, IDN8 with Acc150 = 100% and Acc30 = 80.50%, and IDN12 with Acc150 = 100% and Acc30 = 81.46**%**) are presented.

Table 1, Evaluation parameters for detection of MB/B-points in all subjects IDN1-20 for the complete recording session. Abbreviations: Acc30 – Accuracy (tolerance of ±30 ms), Acc150 – Accuracy (tolerance of ±150 ms), DE30 – Detection Error (tolerance of ±30 ms), DE150 – Detection Error (tolerance of ±150 ms), Missed – the number of detections by alternate metrics, and N – overall number of manual annotations.

| IDN | Acc30 [%] | Acc150 [%] | DE30 [%] | DE150 [%] | Missed | N |
|---|---|---|---|---|---|---|
| 1 | 81.64 | 99.50 | 18.36 | 0.50 | 71 | 1405 |
| 2 | 91.59 | 99.41 | 8.41 | 0.59 | 14 | 1189 |
| 3 | 94.39 | 99.83 | 5.61 | 0.17 | 9 | 1176 |
| 4 | 90.09 | 99.47 | 9.90 | 0.53 | 41 | 949 |
| 5 | 89.41 | 99.67 | 10.59 | 0.33 | 59 | 916 |
| 6 | 54.14 | 99.53 | 45.86 | 0.47 | 738 | 1062 |
| 7 | 75.07 | 99.00 | 24.93 | 1.00 | 40 | 702 |
| 8 | 71.59 | 99.06 | 28.41 | 0.94 | 112 | 1063 |
| 9 | 82.25 | 99.46 | 17.75 | 0.54 | 93 | 924 |
| 10 | 79.89 | 99.82 | 20.11 | 0.18 | 12 | 1119 |
| 11 | 88.67 | 98.92 | 11.33 | 1.08 | 168 | 1112 |
| 12 | 77.74 | 99.69 | 22.26 | 0.30 | 249 | 984 |
| 13 | 93.74 | 98.97 | 6.26 | 1.03 | 77 | 974 |
| 14 | 74.33 | 98.54 | 25.67 | 1.45 | 464 | 1375 |
| 15 | 92.38 | 99.36 | 7.62 | 0.64 | 7 | 932 |
| 16 | 79.09 | 99.60 | 20.90 | 0.40 | 170 | 995 |
| 17 | 89.82 | 99.61 | 10.17 | 0.39 | 47 | 1032 |
| 18 | 94.24 | 99.54 | 5.76 | 0.46 | 46 | 1094 |
| 19 | 90.37 | 99.55 | 9.62 | 0.44 | 90 | 1122 |
| 20 | 81.06 | 99.68 | 18.94 | 0.32 | 54 | 940 |
| All | 83.58± 10.03 | 99.41±0.34 | 16.42±10.03 | 0.59±0.34 | 2561 (sum) | 21065 (sum) |

Table 2, Evaluation parameters for detection of MB/B-points in all subjects IDN1-20 for the baseline recording. Abbreviations: Acc30 – Accuracy (tolerance of ±30 ms), Acc150 – Accuracy (tolerance of ±150 ms), DE30 – Detection Error (tolerance of ±30 ms), DE150 – Detection Error (tolerance of ±150 ms), Missed – the number of detections by alternate metrics, and N – overall number of manual annotations.

| IDN | Acc30 [%] | Acc150 [%] | DE30 [%] | DE150 [%] | Missed | N |
|---|---|---|---|---|---|---|
| 1 | 87.07 | 100 | 12.93 | 0 | 23 | 264 |
| 2 | 97.62 | 100 | 2.38 | 0 | 0 | 211 |
| 3 | 97.30 | 100 | 2.70 | 0 | 5 | 186 |
| 4 | 92.62 | 100 | 7.38 | 0 | 8 | 150 |
| 5 | 90.00 | 100 | 0.10 | 0 | 12 | 151 |
| 6 | 61.45 | 99.44 | 38.55 | 0.56 | 124 | 180 |
| 7 | 89.91 | 100 | 10.08 | 0 | 2 | 120 |
| 8 | 80.50 | 100 | 19.50 | 0 | 16 | 201 |
| 9 | 91.87 | 100 | 8.12 | 0 | 28 | 161 |
| 10 | 73.50 | 100 | 26.50 | 0 | 3 | 201 |
| 11 | 83.43 | 100 | 16.57 | 0 | 40 | 170 |
| 12 | 81.46 | 100 | 18.54 | 0 | 54 | 179 |
| 13 | 97.99 | 100 | 2.01 | 0 | 6 | 150 |
| 14 | 83.09 | 100 | 16.91 | 0 | 51 | 208 |
| 15 | 97.45 | 100 | 2.55 | 0 | 2 | 158 |
| 16 | 93.79 | 99.38 | 6.21 | 0.62 | 7 | 162 |
| 17 | 94.94 | 99.44 | 5.05 | 0.56 | 3 | 179 |
| 18 | 94.90 | 100 | 5.10 | 0 | 8 | 197 |
| 19 | 91.05 | 100 | 8.95 | 0 | 23 | 191 |
| 20 | 78.57 | 100 | 21.43 | 0 | 7 | 169 |
| All | 87.93±9.46 | 99.91±0.21 | 12.07±9.46 | 0.09±0.21 | 422 (sum) | 3588 (sum) |

Table 3, Evaluation parameters for detection of MB/B-point compared with the literature data. Abbreviations: Se – Sensitivity, PP – Positive Predictivity, DE – Detected Error, me – mean error, sd – standard deviation, NA – Not Available, and ICG – Impedance CarioGraphy. Superior parameters corresponding to the new method are highlighted in Bold.

| Reference | Se [%] | PP [%] | DE [%] | me ± sd [%] | N | Subjects |
|---|---|---|---|---|---|---|
| Proposed method (120 s duration, ±150 ms tolerance) | **99.91±0.21** | **100** | / | **0.09±0.21** | 3588 | 20 healthy |
| Proposed method (±150 ms tolerance) | **99.41±0.34** | **100** | / | **0.59±0.34** | **21065** | 20 healthy |
| Proposed method (120 s duration, ±30 ms tolerance) | 87.93±9.46 | **100** | / | 12.07±9.46 | 3588 | 20 healthy |
| Proposed method (±30 ms tolerance) | 83.58±10.03 | **100** | / | 16.42±10.03 | **21065** | 20 healthy |
| Proposed method on data from [2, 30] (±150 ms tolerance) | **96.48±15.80** (99.69±1.01) | **100** | / | 3.52±15.80 (0.30±1.01) | 1833 | 24 healthy (22) |
| Proposed method on data from [2, 30] (±30 ms tolerance) | 66.01±24.87 (67.50±23.06) | **100** | / | 33.99±24.87 (32.50±23.06) | 1833 | 24 healthy (22) |
| [6] for noise-free ICG | 94.4 | 93.9 | 11.7 | / | 545 | 9 healthy and 5 patients |
| [6] for noise-contaminated ICG | 93.4 | 93.0 | 13.6 | / | | |
| [2, 30] (±150 ms tolerance) | 99.04 | 98.13 | 2.92 | 2.03±2.23 | 1920 | 24 healthy |
| [2, 30] (±30 ms tolerance) | 95.30±5.65 | 94.48±6.96 | / | 1.75±0.90 | 1920 | 24 healthy |
| [7] | / | / | / | / | NA | 17 healthy |
| [10] (±30 ms tolerance) | / | / | 1.7 | / | 1517 | 14 healthy and 16 patients |
| [8] | / | 93.1 | 6.9 | / | NA | 40 patients |

## 4 Discussion

Accuracies for MB/B-point detection were higher in cases of baseline ICG recording (Tables 1-2). This is not a surprise as B-point delineation can be sensitive to noise [2]. Expectedly, new detector is also sensitive to the noise. Illustration of the annotated and located MB/B-points in Fig. 3 shows dependencies of proposed peak detector in relation to ICG morphology as reported previously [1, 7, 18]. It should be stated that manual annotation could be erroneous, which is likely the case for the last B-point in IDN12 (Fig. 3). Despite the fact that manual annotation is common for assessment ICG peak detectors in literature [2, 4, 10, 30] it may be subjective for non-distinctive B-points [4]. This probably affected the detection accuracy to some extent as well. Even with all influences, average accuracy of 99.41% for ±150 ms tolerance for complete recording presents satisfactory and superior result.

Missed detections varied a lot: from <1% for IDN3 to ~70% for IDN6 (Table 1). Similar is seen in Table 2. When we compare the sums of missed MB-points and overall number of annotated peaks it appears that alternate B-point detection took part in 12.16% and 11.76% for the complete and baseline recording, respectively. This is not an outstanding difference having in mind variable noise influence. Thus, we may hypothesize that this similar percent of missed MB-point detections originates rather from the ICG morphology than from the artifacts. Interestingly, in [6] Authors obtained similar results as their accuracy was not vastly influenced by noise. An exact comparison with other algorithms cannot be made, but missing B-points were reported to be more than 10% in [18], which is indeed in line with our results.

In general, the comparison with existing methods should be taken with a grain of salt due to the different approaches in B-point delineation, divergent data, and a variety of validation metrics [2, 10] as can be seen for comparison with other methods in Table 3. Still, we summarize results from literature to compare the performance of new algorithm including evaluation of new method for another datasets (Table 3). Detection accuracies and errors showed that the proposed method reaches superior or as good as results in comparison to the existing detectors (Table 3). PP is without any due the most high-ranking parameter of the proposed method. Partly this is result of the MB-point detector dependence on the C-point lineation and partly it is related to the alternate approach for B-point detection in cases transformed ICG segment does not contain MB-point (Fig. 2).

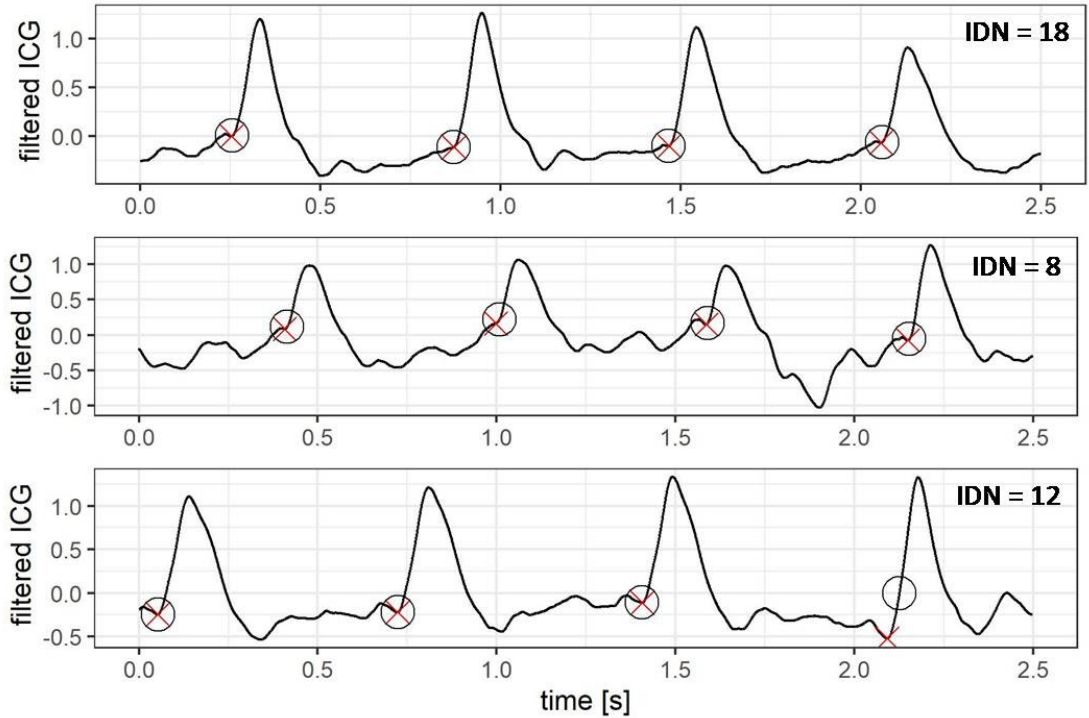

Figure 3, Sample ICG segments in three subjects (IDN = 8, 12, and 18) with manually detected B-points (black circles) and with automatically detected MB/B-points by the new algorithm (red crosses).

The results presented in brackets for data from [2, 30] in Table 3 are given without three recordings as in one case we reached Acc of 0% and hypothesized that the annotations were not saved uniformly. Therefore, results from remaining 42 cases corresponding to 20 healthy subjects are presented. New detector reaches superior results in cases of ±150 ms tolerance and with three discharged recordings. Even when these recordings are included, the results of Acc for new method (96.48%) are comparable with those in [2, 30] (99.04%). The presented dataset in this manuscript is by far the largest one in regards to the number of B-points annotations (Table 3). However, further advancement of the new MB/B-point detector particularly for tolerance of ±30 ms will be welcomed.

### 4.1 Limitations of the Study

We have recognized the following limitations:
1. Future evaluation of the proposed method could be performed in patients [6, 10], as it would be interesting to test the proposed method for different morphology which is of utmost importance for method application in clinical setting.

2. We did not use available signal labeler [32], but we designed our own with the best practice from literature in mind. The comparison of manual annotation procedures exceeds the scope of the manuscript.
3. B-point pattern recognition was not performed and we did not use machine learning algorithms for such purpose [4, 18]. Generally, machine learning presents a promising approach for ICG-based classification. Current trends in biomedical engineering fostered the development of efficient deep learning algorithms for example in ECG-based heart disease detection [33]. As such algorithms do not require feature extraction; their application on ICG signals makes all delineation methods unnecessary. However, deep learning models cannot determine characteristic ICG signatures related to the physiology of the heart (*e.g.*, electromechanical systole intervals). Hence, delineation is required for appropriate feature extraction that is related to the fundamental biological phenomenon. We would argue that for appropriate explanation of causal mechanisms complemented with predictive research questions, a combined approach comprising classical signal analysis (*e.g.*, delineation techniques), machine learning, and deep learning is required.
4. We did not implement signal quality assessment to indicate low quality regions [12]. Future approach could employ such strategy for exact reasoning on method robustness with the quantitative assessment of method robustness to different signal-to-noise ratios.
5. We did not introduce preprocessing adaptation and optimization to enhance method sensitivity to noises and artifacts. One may utilize B-point detectors by excellent results of adaptive Savitzky-Golay filter for noise elimination in ICG [2-3] or in ECG [34-35].

## 5 Conclusions

Transforming ICG with weighted time windows presents a promising technique for MB/B-point detection. The proposed workflow could be directly applied for delineation of other biosignals such as dicrotic notch of the pulse waveform in photopletismogram (PPG) as suggested in [7] or in ECG. We contribute to the body of knowledge in the following ways:

1) We give a detailed workflow for automatic B-point detection that proved superior in PP and in accuracy for tolerance of ±150 ms.
2) Open benchmark data comprising ECG, thoracic impedance ($\Delta Z$), and ICG recorded in 20 university students is available [28]. Besides, we publicly share manually annotated B-points and GNU Octave for manual labeling, as well as R code for B-point detection [28].

Although proposed method is easy to use and can be simply combined with other methods and models for ICG analysis, further improvements and detailed assessment of its vulnerability to noises and artifacts will be welcomed.

## Acknowledgement

This research would not be possible without generous grant to use and openly share the data from Professor Goran Knežević, and Senior Research Associate Ljiljana B. Lazarević, as well as without admirable dedication to conduct measurements by PhD students Olga Dubljević, Bojana Bjegojević, and Nikola Milosavljević from Faculty of Philosophy, Institute of Psychology, and Laboratory for Research of Individual Differences at the University of Belgrade.

## Funding

The Authors were partly funded by the Ministry of Science, Technological Development and Innovation of the Republic of Serbia (Grant No. 451-03-47/2023-01/200103). The Funder was not involved in the study design, collection, analysis, and interpretation of data; in the manuscript preparation; and in the decision to submit the manuscript.

## Institutional Review Board Statement

Institutional Review Board from the Department of Psychology at the University of Belgrade approved the study (No. 2018-19).

## Conflict of Interest Statement

Authors have no competing interests to declare.

## Informed Consent Statement

All participants signed Informed Consents in accordance with the Code of Ethics of the Word Medical Association (Helsinki Declaration).

## Data and Software Availability Statement

For appropriate reproducibility of presented results the R programming and GNU Octave scripts with raw data are publicly available on Zenodo repository [28].

## CRediT Author Statement

**Nadica Miljković**: Methodology, Investigation, Software, Data Curation, Visualization, Writing – Original Draft Preparation. **Tomislav B. Šekara**: Conceptualization, Methodology, Validation, Formal Analysis, Writing – Reviewing and Editing.

## References

[1] Sherwood A, Allen MT, Fahrenberg J, Kelsey RM, Lovallo WR, Van Doornen LJ. Methodological guidelines for impedance cardiography. Psychophysiology. 1990 Jan;27(1):1-23. https://doi.org/10.1111/j.1469-8986.1990.tb02171.x

[2] Pale U, Müller N, Arza A, Atienza D. ReBeatICG: Real-time low-complexity beat-to-beat impedance cardiogram delineation algorithm. In 2021 43rd Annual International Conference of the IEEE Engineering in Medicine & Biology Society (EMBC) 2021 Nov 1 (pp. 5618-5624). IEEE. https://doi.org/10.1109/EMBC46164.2021.9630170

[3] Salah IB, De la Rosa R, Ouni K, Salah RB. Automatic diagnosis of valvular heart diseases by impedance cardiography signal processing. Biomedical Signal Processing and Control. 2020 Mar 1;57:101758. https://doi.org/10.1016/j.bspc.2019.101758

[4] Chabchoub S, Mansouri S, Ben Salah R. Signal processing techniques applied to impedance cardiography ICG signals–A review. Journal of Medical Engineering & Technology. 2022 Jan 13:1-8. https://doi.org/10.1080/03091902.2022.2026508

[5] DeMarzo AP, Lang RM. A new algorithm for improved detection of aortic valve opening by impedance cardiography. In Computers in Cardiology 1996 Sep 8 (pp. 373-376). IEEE. https://doi.org/10.1109/CIC.1996.542551

[6] Naidu SM, Pandey PC, Pandey VK. Automatic detection of characteristic points in impedance cardiogram. In 2011 Computing in Cardiology 2011 Sep 18 (pp. 497-500). IEEE. https://ieeexplore.ieee.org/abstract/document/6164611

[7] Carvalho P, Paiva RP, Henriques J, Antunes M, Quintal I, Muehlsteff J. Robust characteristic points for ICG-definition and comparative analysis. In BIOSIGNALS 2011 Jan 26 (pp. 161-168).

[8] Salah IB, Ouni K. Denoising of the impedance cardiographie signal (ICG) for a best detection of the characteristic points. In 2017 2nd International Conference on Bio-engineering for Smart Technologies (BioSMART) 2017 (pp. 1-4). IEEE. https://doi.org/10.1109/BIOSMART.2017.8095347

[9] Shyu LY, Lin YS, Liu CP, Hu WC. The detection of impedance cardiogram characteristic points using wavelet transform. Computers in Biology and Medicine. 2004 Mar 1;34(2):165-75. https://doi.org/10.1016/S0010-4825(03)00040-4

[10] Bagal UR, Pandey PC, Naidu SM, Hardas SP. Detection of opening and closing of the aortic valve using impedance cardiography and its validation by echocardiography. Biomedical Physics & Engineering Express. 2017 Nov 27;4(1):015012. https://doi.org/10.1088/2057-1976/aa8bf5

[11] Cacioppo JT, Tassinary LG, Berntson G, editors. Handbook of psychophysiology. Cambridge University Press; 2007 Mar 5.

[12] Nabian M, Yin Y, Wormwood J, Quigley KS, Barrett LF, Ostadabbas S. An open-source feature extraction tool for the analysis of peripheral physiological data. IEEE Journal of Translational Engineering in Health and Medicine. 2018 Oct 26;6:1-1. https://doi.org/10.1109/JTEHM.2018.2878000

[13] Kauppinen P, Hyttinen J, Laarne P, Malmivuo J. A software implementation for detailed volume conductor modelling in electrophysiology using finite difference method. Computer Methods and Programs in Biomedicine. 1999 Feb 1;58(2):191-203. https://doi.org/10.1016/S0169-2607(98)00084-4

[14] Antić M, Popović NB, Milosavljević N, Dubljević O, Bjegojević B, Miljković N. CardioPRINT: Individual features hidden in electrocardiogram and impedance-cardiogram. Empirical Studies in Psychology. 2020 Oct 15:13.

[15] Badeli V, Jafarinia A, Melito GM, Müller TS, Reinbacher-Köstinger A, Hochrainer T, Brenn G, Ellermann K, Biro O, Kaltenbacher M. Monitoring of false lumen thrombosis in type B aortic dissection by impedance cardiography–A multiphysics simulation study. International Journal for Numerical Methods in Biomedical Engineering. 2023 Feb;39(2):e3669. https://doi.org/10.1002/cnm.3669

[16] Kubicek WG, Patterson RP, Witsoe DA. Impedance cardiography as a noninvasive method of monitoring cardiac function and other parameters of the cardiovascular system. Annals of the New York Academy of Sciences. 1970 Jul;170(2):724-32. https://doi.org/10.1111/j.1749-6632.1970.tb17735.x

[17] Atzler E, Lehmann G. Über ein neues verfahren zur darstellung der herztätigkeit (Dielektrographie). Arbeitsphysiologie. 1932 May 5;5(6):636-80. https://doi.org/10.1007/BF02008706

[18] Benouar S, Hafid A, Attari M, Kedir-Talha M, Seoane F. Systematic variability in ICG recordings results in ICG complex subtypes–steps towards the enhancement of ICG characterization.

Journal of Electrical Bioimpedance. 2018 Jan 1;9(1):72-82. https://doi.org/10.2478/joeb-2018-0012
[19] Hung PD, Dan CQ, Hai VD. A method for suppressing respiratory noise in impedance cardiography and comprehensive assessment of noise reduction performance. Journal of Medical Engineering & Technology. 2022 Feb 17;46(2):116-28. https://doi.org/10.1080/03091902.2021.2007304
[20] Team RC. R: A language and environment for statistical computing. 2013
[21] Wickham H, Francois R, Henry L. Müller K. dplyr: A grammar of data manipulation. R package version 0.4. 3. R Found Stat Comput Vienna https. CRAN R-project org/package= dplyr. 2015
[22] Developers. Signal: Signal processing. http://r-forge.r-project.org/projects/signal. 2013
[23] Hadley W. Ggplot2: Elegant graphics for data analysis. Springer; 2016.
[24] Borchers HW. Pracma: Practical numerical math functions. R package version. 2019;2(9):519.
[25] Bjegojević B, Milosavljević N, Dubljević O, Purić D, Knežević G. In pursuit of objectivity: Physiological measures as a means of emotion induction procedure validation. Empirical Studies in Psychology. 2020 Oct 15:17.
[26] Boljanić T, Miljković N, Lazarević LB, Knežević G, Milašinović G. Relationship between electrocardiogram-based features and personality traits: Machine learning approach. Annals of Noninvasive Electrocardiology. 2022 Jan;27(1):e12919. https://doi.org/10.1111/anec.12919
[27] Pan J, Tompkins WJ. A real-time QRS detection algorithm. IEEE Transactions on Biomedical Engineering. 1985 Mar;(3):230-6. https://doi.org/10.1109/TBME.1985.325532
[28] [dataset and software] Miljković N, Šekara TB. Software for detection of B-point in impedance cardiogram with dataset consisting of 20 healthy participants 2023. https://doi.org/10.5281/zenodo.6813716
[29] Eaton JW, Bateman D, Hauberg S, Wehbring R. GNU Octave version 4.0. 0 manual: A high-level interactive language for numerical computations. 2015. URL http://www.gnu.org/software/octave/doc/interpreter. 2019;8:13.
[30] [dataset] Pale U, Meier D, Müller O, Valdes AA, Alonso DA. ReBeatICG database 2021. https://doi.org/10.5281/zenodo.4725433.
[31] Bengtsson H. R.matlab: Read and Write MAT Files and Call MATLAB from Within R. R package version 3.6.1-9000, 2017. https://github.com/HenrikBengtsson/R.matlab
[32] [software] Zanoli S, Teijeiro Campo T, Atienza Alonso D. Physiological signals labeler 2021. https://doi.org/10.5281/zenodo.4724843.
[33] Papageorgiou VE, Zegkos T, Efthimiadis G, Tsaklidis G. Analysis of digitalized ECG signals based on artificial intelligence and spectral analysis methods specialized in ARVC. International Journal for Numerical Methods in Biomedical Engineering. 2022 Nov;38(11):e3644. https://doi.org/10.1002/cnm.36444
[34] Popović NB, Miljković N, Šekara TB. Electrogastrogram and electrocardiogram interference: Application of fractional order calculus and Savitzky-Golay filter for biosignals segregation. In 2020 19[th] International Symposium INFOTEH-Jahorina 2020 Mar 18 (pp. 1-5). IEEE. https://doi.org/10.1109/INFOTEH48170.2020.9066278
[35] Zadeh AE, Khazaee A, Ranaee V. Classification of the electrocardiogram signals using supervised classifiers and efficient features. Computer Methods and Programs in Biomedicine. 2010 Aug 1;99(2):179-94. https://doi.org/10.1016/j.cmpb.2010.04.013